# Elementary School Students' and Teachers' Perceptions Towards Creative Mathematical Writing with Generative AI

Yukyeong Song [a]

Jinhee Kim [b*]

Wanli Xing [a]

Zifeng Liu [a]

Chenglu Li [c]

Hyunju Oh [a]

[a] *School of Teaching and Learning, College of Education, University of Florida, FL, United States*

[b] *Department of STEM Education and Professional Studies, Old Dominion University, Norfolk, VA, United States*

[c] *Educational Psychology, College of Education, The University of Utah, Salt Lake City, UT, United States*

\* Corresponding Author Info:

Jinhee Kim

4301 Hampton Blvd, Room 2333, Norfolk, VA 23529

Tel: +1 757 683 5163   email: jhkim@odu.edu

**Abstract**

While mathematical creative writing can potentially engage students in expressing mathematical ideas in an imaginative way, some elementary school-age students struggle in this process. Generative AI (GenAI) offers possibilities for supporting creative writing activities, such as providing story generation. However, the design of GenAI-powered learning technologies requires careful consideration of the technology reception in the actual classrooms. This study explores students' and teachers' perceptions of creative mathematical writing with the developed GenAI-powered technology. The study adopted a qualitative thematic analysis of the interviews, triangulated with open-ended survey responses and classroom observation of 79 elementary school students, resulting in six themes and 19 subthemes. This study contributes by investigating the lived experience of GenAI-supported learning and the design considerations for GenAI-powered learning technologies and instructions.

*Keywords*: Generative AI in education; AI-supported math learning; mathematical creative writing; student perception; teacher perception

# I. Introduction

Creative writing is an expression of fiction or non-fiction stories that encompass narratives, characters, emotional evocation, and thought provocation (Wise & Luyn, 2020). Creative writing is often used as a pedagogical strategy, allowing students to understand, construct, synthesize, and elaborate on learned knowledge with their cultural inspirations (Bloom, 1956; Wills, 1993). The importance and potential of creative writing's role in supporting learning have been increasingly discussed (Colonnese et al., 2018; Firmender et al., 2017). Elementary Mathematical Writing Task Force has emphasized mathematical writing suggesting mathematical creativity as a goal of mathematical communication and reasoning (Firmender et al., 2017). Creative writing in mathematics encourages students to document their original and creative stories that convey mathematical concepts, ideas, problems, and solutions, enhancing fluent and flexible thinking skills (Casa et al., 2016). However, young students or students with special needs have difficulty engaging in such creative writing activities, especially when lacking supporting tools and resources (Colonnese et al., 2018).

Generative artificial intelligence (GenAI) brings new and varied opportunities to support students' creative writing activities in mathematics. For instance, students could collaborate with AI to ignite their creative writing journey creating a less intimidating experience than starting from scratch. AI could provide an initial spark to kickstart story generation through customized prompts reflecting individual learners' cultural backgrounds, interests, and preferences. In this process, GenAI-powered agents could provide scaffolding for students to use mathematical language, encouraging them to expand their stories through creativity and imagination (Zhang et al, 2024). Storytelling captures and maintains the learner's attention, increasing interest and motivation (Hava, 2021), and supporting math language learning and story creation performance (Zhang et al., 2024).

Although the diverse roles of AI in creating mathematical writing and its benefits in the K-12 learning context have been explored, designing and implementing effective AI-assisted creative mathematics writing solutions face challenges. These challenges include overreliance on AI-generated content (Kim et al., 2024a; Chan & Hu, 2023), reduced opportunities for social learning and peer/teacher feedback (Zimmerman et al., 2023; Guilherme, 2019), and exposure to inappropriate content (Kim et al., 2024a). To better address such limitations, maximize the pedagogical benefits of creative mathematical writing, and effectively take advantage of GenAI technologies, it is important to account for both students' and teachers' perspectives of the lived reality of the classroom and their needs on the ground to create more constructive and meaningful interactions among students, teachers, and the AI system (Heilporn et al., 2021). Understanding how students and teachers perceive what they do, construe, and evaluate their learning and teaching with technology and why they choose to act in particular ways in specific circumstances offers important insights. These gained insights could provide a future direction for the design of AI technologies in education.

This study explores students' and teachers' perceptions about creative mathematical writing with GenAI earned from an empirical classroom implementation involving 79 5th-grade students and six teachers. Students and teachers engaged in creative mathematical writing activities supported by GenAI-powered learning technology, [ANONYMOUS], developed by our research team. After the four class periods of experiencing [ANONYMOUS], a subset of students and all six teachers were interviewed. This study is especially interested in identifying the opportunities and challenges of using GenAI to support creative mathematical writing, as perceived by students and teachers. Two research questions were set to guide this study.

- RQ1. What are the educational benefits of creative mathematical writing with GenAI as perceived by students and teachers?
- RQ2. What are the challenges of creative mathematical writing with GenAI as perceived by students and teachers?

## II. Literature review

### 2.1. Creative Mathematical Writing

Traditional classroom settings often rely on the "chalk-and-talk" teaching environment. In mathematics education, these settings tend to prioritize accuracy practice and finding the "right" answer over fostering deep understanding (Freire, 2020), exploring underlying concepts, or encouraging creative thinking. Moreover, in a traditional math classroom, most students, particularly low-performing ones, have limited opportunities to act as designers or tutors or to engage in the creative process of mathematics (Dewsbury, 2020), which in turn negatively impacts academic performance and self-regulated learning (Kim et al., 2024b), broadening disparities among students (Byers et al., 2018).

In line with this, creative mathematical writing has been proposed as an effective learner-centered pedagogy. Mann (2006) also described mathematical creative writing as the essence of mathematics that engages students in advanced mathematical thinking and learning. Engaging students in creative mathematical writing could create immersive and contextual learning scenarios where students can act as designers and owners of their stories (Kim & Lee, 2023). In addition, these activities stimulate deeper cognitive processing and retention, helping students understand complex thoughts and ideas while having fun (Kim & Cho, 2023; Zhang et al., 2024). For instance, student-generated real-life math stories can facilitate the identification and representation of mathematical relationships, making abstract concepts more tangible and understandable (Overholt et al., 2008).

*2.2. Educational Benefits and Challenges of GenAI-supported Writing*

GenAI generates new content (e.g., texts, images, audio) similar to existing data by learning the features and patterns of that data (Su & Yang, 2023). Research demonstrates that using GenAI (e.g., text-to-image AI) significantly boosts human creative productivity and increases the value of the content (Zhou & Lee, 2024). ChatGPT usage in writing can significantly raise writing productivity by generating text content (Kim et al., 2024a). Its capabilities in improving language and proofreading are well-established (Shopovski, 2024), making AI-generated content trustable with appropriate monitoring and revision (Kim et al., 2024a).

Recently, attempts have been made to integrate GenAI into educational scenarios to support creative writing activities. For example, the educational software, "Plotagon," allows students to create AI-generated visualizations for stories, blending creative writing with language learning (Guzmán Gámez & Moreno Cuellar, 2019). In mathematics education, Zhang et al. (2024) introduced "Mathemyths," a joint storytelling agent that co-creates stories with children, integrating mathematical terms into the evolving narrative. A user study involving 35 children aged 4-8 years indicated that interactions with "Mathemyths" improved learning outcomes for mathematical language that were comparable to those achieved when co-creating stories with a human partner. GenAI can also enrich narrative development and offer new perspectives in storytelling, and using AI as a pedagogical tool can provide qualitative educational benefits, encouraging further research into AI-assisted learning strategies in education (Wafa et al., 2024).

Despite the educational potential of GenAI-supported writing, it faces several challenges. One major concern is that young learners may become overly reliant on AI-generated information, which could restrict their creativity and originality (Chan & Hu, 2023; Kim et al., 2024a). This dependency can limit opportunities for social learning and feedback from teachers and peers, which are crucial for cognitive and social development (Zimmerman et al., 2023; Guilherme, 2019). AI-generated stories might also unintentionally contain inappropriate, biased, or offensive content, potentially exposing learners to harmful ideas. Ensuring learners' safety when interacting with AI systems for storytelling requires vigilance and ongoing improvements in AI models (Kim & Cho, 2023; Kim et al., 2024a). Another challenge is maintaining educational standards and alignment with learning objectives. AI-generated content must be tailored to meet curricular requirements and foster meaningful learning experiences. Effective integration of AI into teaching and learning requires thoughtful instructional design to balance technology use with human-centric teaching (Weng & Chiu, 2023). Furthermore, fostering relationships between students and their teachers or peers enhances students' sense of ownership and agency in AI-supported learning (Wentzel, 1998). These relationships increase engagement and motivation, leading to more effective and self-directed learning experiences.

## III. Research methods

*3.1. Research Context and Participant*

We conducted classroom studies to implement our GenAI-powered learning technology in six classrooms led by six elementary school teachers with 79 5th-grade students across two United States (US) schools. One school, located on the West Coast of the US, involved four class sessions, each session lasting 45-60 minutes, covering fraction division in math with three teachers and fifty-seven students. The other school, located in the Southeast US, involved two classes spanning eight sessions, each lasting 35 minutes, covering multiplications of fractions in math with three teachers and 22 students. Teacher participants included five females and one male. The demographic breakdowns of the students included 50 girls, 56 boys, one non-binary, and five chose not to say. In addition, 36 students identified themselves as White, 10 as Asian, eight as Black/African American, three as Native American/Alaska Native/First Nations, and two as Native Hawaiian/Pacific Islander. Sixteen students selected 'self-specify,' and answers couldn't be categorized into any of the provided options (e.g., Mexican; Brazilian), while 10 chose not to specify. As participants were allowed to select more than one category, the total count of selections exceeded the number of participants. This study received ethical approval from the [author's university]'s Institutional Review Board (IRB No. 202300714) and informed consent forms from all participants.

### *3.2. Instruction and Learning Technology*

We followed a three-step instruction to support students' creative mathematical writing, conceptualized as Ask, Represent, and Transform. First, in the *Ask* step, students are asked questions related to the math stories provided to them, break down the math stories, and deeply understand the math ideas behind the stories. For example, there is a story titled "Garden Traditions with Abuela," as follows:

> "Sofia was excited for Saturday. She planned to help her Abuela (grandmother) plant some zucchinis and tomatillos in her garden. … When she arrived at the garden, her Abuela had already plated ¼ of a row of zucchinis and suggested that Sofia wait until she finished planting zucchinis and help her with planting tomatillos. If it took Abuela 6 minutes to plant ¼ of a row of zucchinis, how long will it take her to plant 1 row of zucchinis?"

After reading this story, students are prompted to answer the following questions: *"What is the story about?" "What else do you wonder about the story?"* and *"Describe or select the math ideas in the story"* (refer to Figure 1. (a)).

In the *Represent* step, students are encouraged to model their mental representations of the possible solutions, utilizing different visual tools provided by the platform. They are provided with an empty canvas they can fill with different shapes, symbols, and texts (refer to **Figure** 1. (b)).

Finally, in the *Transform* step, the GenAI technology supports students in drafting, modifying, and creating their culturally relevant math stories. In this stage, students take on the role of a director by providing story ideas and goals (i.e., math-related goals or a math equation the story aims to convey), character (i.e., the main character of the story), settings (i.e., the major plot for the story), and event (i.e., the basic outline of the story including what is happening in the story related to math ideas), as shown in Figure 1 (c)-left. In this example, the student wrote the goal to be "*⅗ X 15 = 9,*" the character to be "*Kimberly,*" and the setting to be "*Pizza shop.*" The event and math problem are written as "*Kimberly loves to eat pizza. She ate ⅗ of a giant pizza. She is full but she wants to eat the rest of the pizza.*" Based on the student's input, GenAI created a story draft, as shown on the right-hand side of **Figure** 1 (c):

> "Kimberly, a big pizza lover, goes to her favorite pizza shot after school each day. One day, Kimberly felt extra hungry. A whole, delicious pizza was in front of her. (...) She had eaten

⅗ of the pizza. But, she felt quite full already. Kimberly wondered, how much more of the pizza does she need to eat to finish the whole pizza?"

After that, students were given a checklist to evaluate the AI-generated stories. The checklist includes questions like *"Does the story make sense?" "Is the story interesting?"* and *"Does the story include the math equation you intended?"*.

We integrated multimodal generative AI by leveraging OpenAI's GPT-4 and DALL·E 3, renowned for their capabilities in text and image synthesis (Betker et al., 2023). In our application, students' input populate a structured prompt designed for math story creation. This prompt has undergone red teaming (Perez et al., 2022) to evaluate its effectiveness in producing readable, accurate, and safe text. Enhancements to text generation include the use of active prompting techniques (Diao et al., 2023), which draw on few-shot learning principles. Active prompting allows the AI to dynamically adjust its example selection for generation guidance to optimize performance and relevance based on the specific contexts (e.g., students' goals). Our platform utilizes a curated database of primary school-level mathematical stories crafted and reviewed by educational experts. The application selects two stories that best match the student's input, the mathematical topic, and the grade level to guide the generation process. After generating a story with GPT-4, DALL·E 3's API activates to create corresponding images.

**Figure** 1. A screenshot of the interface for [ANONYMOUS]

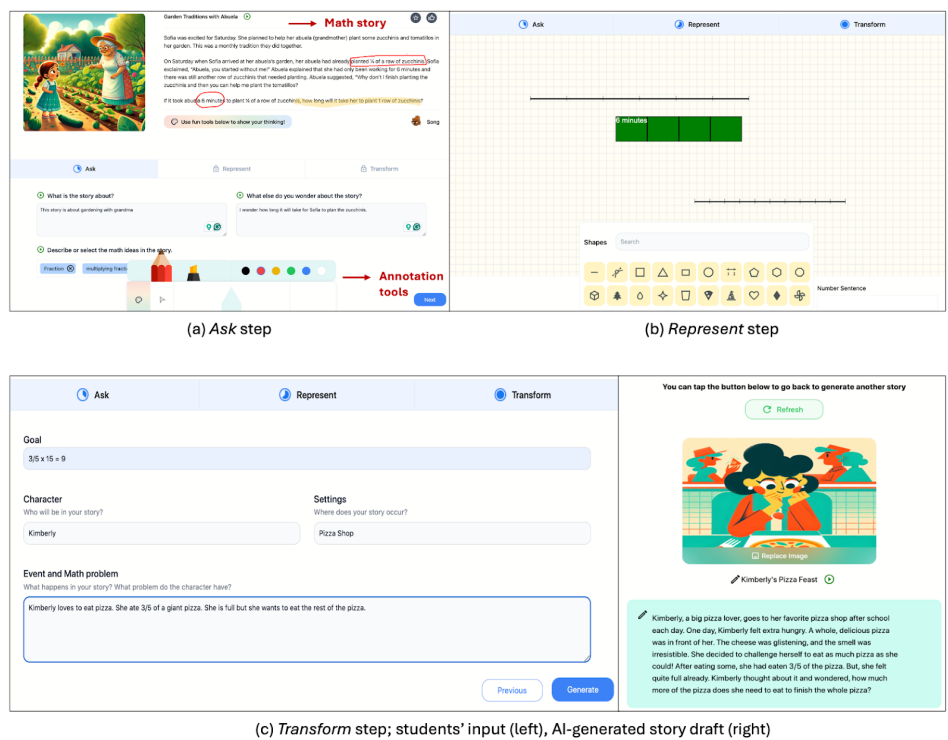

(a) *Ask* step

(b) *Represent* step

(c) *Transform* step; students' input (left), AI-generated story draft (right)

*3.3. Data Collection*

We collected qualitative data from students and teachers during and after the classroom study to identify perceptions of elementary school students and teachers on GenAI-supported creative mathematical writing. We conducted semi-structured focus group interviews with a subset of invited students to learn about students' perceptions of using GenAI in creative mathematical writing. Researchers and teachers discussed and decided which students to invite based on the student's participation in the study and the class activities, communication skills, and representation of diverse demographic backgrounds (e.g., diverse gender and race/ethnicity). A total of 10 students were divided into two groups of five and participated in 30-minute-long in-person interviews. Each focus group was led by a graduate researcher, and the whole interview was voice-recorded and transcribed using an automatic transcription service, Otter.ai. Humans reviewed and corrected the transcriptions based on the audio files afterward. After the classroom studies, we also conducted one-on-one interviews with the six participating teachers to learn about their perceptions through a web conferencing application, Zoom, which were also voice-recorded and transcribed. While the interview data were used as a main data source to draw themes, we also utilized other supplementary data sources to triangulate what we found in the interviews. First, we collected students' responses to the open-ended survey questionnaires, which asked, "*What did you enjoy most about writing math stories with AI?*" and "*How can we improve your experience of writing math stories with AI?*" Second, researchers and teachers took observation notes during the classroom implementation.

*3.4. Data Analysis*

We employed a mixed approach of inductive and deductive thematic analysis (Braun & Clarke, 2006) to analyze the collected data from participants. First, two researchers in a team reviewed the interview transcription data and independently conducted a deductive thematic analysis, applying their expertise to classify data into patterned themes (Glesne, 2016). Following this, an inductive analysis was conducted to discover potential codes and themes not initially identified. The generated themes were then thoroughly reviewed and triangulated with the survey and observation data. In addition, researchers combined themes or divided themes into sub-themes by clarifying the meaning of codes with exemplary quotes based on the participants' utterances from the interview. This iterative process continued until two researchers reached a consensus on each theme.

# IV. Findings

Thematic analysis of the interviews with students and teachers resulted in six themes and nineteen sub-themes, including four themes with 12 sub-themes on the benefits of creative mathematical writing with GenAI (RQ1) and two themes with seven sub-themes on the challenges of GenAI-assisted creative writing (RQ2). **Table** 1 summarizes the themes, sub-themes, and exemplary quotes for each RQ.

**Table** 1. Summary of emergent themes

| Category | Themes | Sub-themes | Exemplary quotes |
|---|---|---|---|
| Benefits of creative mathematical writing with GenAI (RQ1) | T1. Fostering creativity in Math | Promoting divergent thinking | I didn't have to worry about messing it up or taking a long time to create different stories because AI has unlimited chances for me to create stories quickly. So, I tried different math ideas with different characters, settings, and plots (S1). |
| | | Promoting convergent thinking | I liked how AI gave a story and corrected my mistakes. When I write it, I just write whatever I wanted to add and the AI corrects me, added more detail, and made it into one single story (S2). |
| | | Promoting original creations | Based on what AI provided me, I made it really unique; I added fun pictures and made the story more interesting. I did not find similar stories from other in my class (S9). |
| | | Promoting flexible thinking | What I really liked was adding more to the story that I wrote, which I think opened my eyes to other ways to solve problems (S4). |
| | | | This AI system helped them to understand the problem from different angles. Students have to think of multiple strategies that match or solve using numbers (T3). |
| | T2. Fostering mathematical understanding | Understanding the problem | I read the story about a problem, answered questions, and drew a picture about it. Then I realized what was the real problem (S4). |
| | | | I think it really encourages the students to break down the problem and understand all the different pieces of the problem, which students skip over a lot of that part if you don't make them do it with you (T1). |
| | | Mathematical knowledge representation | The canvas and the tools helped me figure out the numbers, lines, and shapes to make the word problems simpler and easier to understand (S5). |
| | | Uncovering misconception | Students often solve numerical problems without completely understanding the underlying mathematical concept. But this AI-supported practice offers students the story or scenario of their interest and allows them to verbalize the misconceptions explicitly as they write the story and communicate with the system (T3). |
| | T3. Promoting AI literacy | Utilizing prompt engineering strategies for AI | I kept trying to make the story better. My first story had many problems. I went back to my direction and changed my requests. I realized AI needs very specific requests to make a good story (S3). |
| | | Evaluating AI's strengths and weaknesses | It helped me understand that AI is not going to be perfect. You can always change your story, and it's not supposed to be perfect. It can be any way you want it to be (S4). I can create new stories, like adding the main content as a human being. AI can add details and put it together into an interesting story quicker than I can (S5). |
| | T4. Improving affective domain | Promoting interest in learning | I always thought math was the most boring subject, but it was so much fun this time so I wanted to do it after school. Working with AI was so much fun (S1). |
| | | | They seemed so interested in the topics of the class with the creation of the problem which they already feel familiar in their life together with the AI system (T1). |
| | | Sparking curiosity in learning | From connecting my story to the math problems and reading the stories that the AI generates for me, I kept posing questions like why and how they could relate. (S2) |
| | | | Students stay curious about understanding the problems, exploring and discovering different strategies to solve problems as they interact with the system to develop the story and integrate mathematical argumentation (T1) |
| | | Developing ownership of learning | Oh, I can create my own math problems and maybe my classmates will read it. So, I have to make sure to understand and solve them for myself first (S2). |
| | | | They usually work on the math problems that someone has already created but students now view themselves as the author of the math problems, making up their own story problems, and they feel responsible for the way they engage with course content and demonstrate their learning (T3). |
| Barriers to AI-supported story-based math learning (RQ2) | T5. Students-related | Lack of critical evaluation of the AI-generated contents | The AI's story was not perfect, but I didn't change anything. I know it's not perfect but I didn't know which part I could improve (S3). |
| | | Problem generation competence | When I had to generate stories, I had to make sure that the story made sense, so that the AI could also understand it and help me solve the problem. But I didn't know what to present at first (S6). |
| | T6. AI-related | Lack of room for improvement | To me, the AI's story was too perfect. When I first looked at it, it was entertaining and made sense. I don't like how the AI is too perfect because I don't want the story to be exactly kind of what I wrote down (S3). |
| | | Lack of pedagogical skills | I wish the system could give me feedback once I submit my story, and then it tells me what I did wrong or how to improve it. Sometimes, I needed feedback from my teacher. I wish I can press a button in the system to send a message to my teacher (S6). |
| | | | When students need help, they can ask, raise their hands, and speak up during the class. But while they were working on the problems with the AI system, students sought help and assistance when they encountered it. As the system couldn't respond promptly, they complained and came to me for advice (T5) |
| | | Deprivation of autonomous learning opportunities | I wanted to create my own story without the AI's help as well. I think there should be a button saying "make your own story" where I can create my story from scratch too (S7). |
| | | Lack of social connection | I wanted to see what kinds of stories my friends created. I also wanted to show my work to the class. My teacher showed us some of my classmates' works on the big screen, but I wish I could go into others' works freely in the system (S4). |
| | | | Kids are interested in how others perform as well. They wanted to see how others worked on the story-making and what math problems they came up with. And they also wanted to have the voting stuff and show something like 'Like' for those good ones. But the system didn't allow them to help their friends, and be engaged with interactions with their peers (T5). |
| | | Lack of multiple means of communication | If I could use the microphone on the computer and the system could automatically write down what I was saying, it would be so much more helpful to discuss the story and math problems (S7). |
| | | | Some students cannot type fast and that's where the speech-to-text would be helpful so they can say all of their ideas easily (T6). |

*4.1. Students' and teachers' perceived benefits of creative mathematical writing with GenAI*

*4.1.1. Fostering creativity in math*

Creative mathematical writing with GenAI helps foster creativity in math. First, the support of GenAI was found to promote students' divergent thinking in creative writing. Divergent thinking means producing as many ideas as possible for an open-ended problem without worrying about judgment (Razumnikova, 2020). Students felt less pressure on the writing task as GenAI provided the draft for them to realize their initial ideas without an intense commitment, time, and energy consumption, which helped them make multiple attempts to create divergent stories. On the other hand, GenAI also provides support for convergent thinking. Convergent thinking refers to refining ideas into one (correct or most appropriate) solution (Razumnikova, 2020). In our study, GenAI provided a single story as a draft by putting together students' divergent ideations and trials. This process gives students a hint on how to correct, expand, and improve their writing, focusing on important mathematical concepts. Third, GenAI can promote students' original creation of mathematical stories. GenAI allows students to input their own ideas as prompts and create original stories for everyone. Students appreciate the originality GenAI provided and incorporate it into their original stories. Lastly, GenAI promotes students' fluent and flexible thinking in mathematical problem-solving. Students infused mathematical ideas into the story and provided possible solutions to the story. As the stories generated by AI are not as structured as students' written stories, students could come up with multiple math problems and solutions to one story. Contrary to the previous literature and common concerns about GenAI's risk of reducing creativity (Shidiq, 2023), this theme suggests that GenAI supports various aspects of creativity, such as divergent thinking, convergent thinking, originality, and flexibility. This finding supports the current discussion on the potential of GenAI in enhancing students' creative writing in mathematics by helping with character development, setting description, and narrative development (Nouari et al., 2024).

*4.1.2 Fostering mathematical understanding*

Creative mathematical writing with GenAI helps students to foster mathematical understanding. First, both students and teachers reported that their learning experience enhanced students' abilities to understand the problem. Understanding problems is considered the first step in mathematical problem-solving (Balım, 2009). As emphasized by the teacher's quote in **Table** 1, the critical step of *"breaking down the problem"* is often omitted in the classroom unless students are accompanied by the teachers or facilitators throughout this process. One of the preparation steps before generating stories (i.e., *Ask* step) encouraged the students to ask questions about the story and helped them understand the details. This process helps students confirm their understanding of mathematical concepts, identify areas where they need further clarification, build personal meanings within the story, and engage more actively with the learning content (Bandura, 2001; Linnenbrink-Garcia et al., 2013). Second, creative math stories with GenAI helped students with mathematical knowledge representation. The *Represent* step in our curriculum allows students to visually represent their mathematical ideas and solutions, making abstract mathematical concepts more tangible and understandable (Overholt et al., 2008). Lastly, teachers emphasized that creative mathematical writing was found to be helpful in uncovering students' misconceptions about math. Compared to traditional math problem-solving practices, such as multiple choice or short response questions, story-based math problem-solving includes symbolizing students' (mis) conceptions and problem-solving processes through writing (McGuire et al., 2008). Therefore, teachers could investigate students' mental modeling of problem-solving and find out conceptual changes or misconceptions that students have.

*4.1.3 Promoting AI literacy*

Working on creative mathematical writing with GenAI was illustrated to promote students' AI literacy. AI literacy refers to competencies needed to live, work, and communicate with AI technologies (Long & Magerko, 2020). In this study, we found that students utilized different prompt engineering strategies, which is an important competency needed to work with GenAI (Knoth et al., 2024). While collaboratively writing math stories with AI, students naturally learned the importance of prompt engineering and internalized better prompt engineering strategies by trial and error. This finding resonates with Woo and colleagues' (2022) finding that secondary students tried different input units (i.e., words, sentences, or paragraphs) in their creative writing work with AI. This activity is an essential part of AI literacy in that it promotes better use of AI and valuable learning of how to evaluate AI's strengths and weaknesses (Long & Magerko, 2020). By recognizing that AI is not always perfect, they, in turn, realize their strengths as humans and acknowledge that they can control the AI's performance.

*4.1.4. Improving affective learning domain*

Students and teachers illustrated that creative mathematical writing with GenAI improves learners' affective learning domains. First, creative mathematical writing with GenAI made learning more interesting and enjoyable. Students found learning fun and showed enthusiasm for learning math as they created stories in collaboration with AI on topics related to their lives (e.g., video games, collaborative schoolwork, or gardening with grandma). Students reflected on their areas of interest and related them to further hypothesize, interpret, and solve mathematical problems. Similarly, teachers reported that learning through topics with personal meanings and creating culturally relevant stories through collaboration with the AI system increased students' attention, focus, and interest in learning.

Second, creative mathematical writing with GenAI sparked their curiosity in learning. GenAI's responsive characteristics, which receive students' inputs and generate personalized stories for individual students, piqued students' curiosity about what kinds of stories they can create with AI support. Additionally, students were encouraged to pursue their curiosity about the problems and explore different solution strategies by asking and answering questions about the problems and utilizing the knowledge they gained to interact with the AI system and create their own stories.

Third, creative mathematical writing with GenAI was found to develop students' feelings of ownership in learning. Compared to traditional math classes, where students are asked to solve existing math problems, participating students were considered authors creating math stories and problems that could potentially be delivered to their peers to solve. The AI system supports this ownership by requiring them to set the story's main elements and the math ideas behind it. Moreover, students were asked to critically evaluate and edit or even rewrite the stories generated by AI. On top of this, students' expectations that their stories could be shared with their peers and teachers made them feel more responsible for their learning and more motivated to understand the concepts with an enhanced sense of ownership as authors of the problems (Hava, 2021).

### *4.2. Students' and teachers' perceived challenges to mathematical creative writing with GenAI*

*4.2.1 Students-related*

First, students' lack of critical evaluation of the AI-generated content was pointed out to be a challenge. While AI generates the content, it is paramount for humans to critically evaluate the AI-generated content (Kim et al., 2024a). Studies raised concerns about students' overreliance on AI-generated content and accepting it without critical examination, which could yield inaccurate, unethical, or low-quality writing (Chan & Hu, 2023). However, even with high-performing AI that can provide accurate and high-quality content, students' examination of the AI-generated content and editing and restructuring of the content provides a valuable learning opportunity for them to construct domain knowledge (Barrett & Pack, 2023). Therefore, it is important to provide pedagogical support to promote students' critical examination of AI-generated content and meaningful collaboration with AI.

Second, students perceived difficulty articulating and formulating coherent stories aligning with the AI's capabilities and learning objectives. Students were not proficient in effectively leveraging AI's strengths or avoiding its weaknesses when generating math stories. Although GenAI provided age-appropriate technologies for students to draft, modify, and refine their math stories, this process could still be challenging for some students, as they still need to create new visual assets associated with the story and generate original mathematical ideas and story plots.

*4.2.2 AI-related*

Students and teachers expressed many AI-related challenges. First, students expressed that GenAI did not provide enough room for improvement, saying, "AI's story was too perfect." This is an interesting finding in the educational context, where the purpose of using AI is to support student's cognitive development rather than to merely produce high-quality writing outcomes (Barrett & Pack, 2023). If the AI performs perfectly in the educational context and provides perfectly complete outcomes, students would be deprived of learning opportunities they can earn by critically evaluating, editing, and improving the content. This implies an important insight into the considerate adjustment of performance levels of GenAI in education.

Second, it was found that the AI system lacks the necessary pedagogical skills, such as providing adaptive and formative feedback. In addition, students expressed that the stories and learning materials failed to adapt to individual students' levels ("*It was too easy for me. It will be different for everyone, but there are not many options for us*," S6). Students expected the AI system to be a content generator for math stories and a pedagogical agent that could provide adaptive and personalized support throughout the learning process. Teachers specifically emphasized the importance of the AI system's function in providing adaptive and formative feedback to the students, as it could provide preliminary instant feedback before teachers.

Third, it was found that creative mathematical writing with GenAI deprived students of autonomous learning opportunities. While GenAI was introduced as a learning support for those students who need a kick-start in writing, some students expressed that they wanted to do it by themselves as well. This could be related to students' intrinsic motivation, which is stimulated when they have more autonomy in their learning (Kim & Cho, 2023). This finding provokes researchers to reevaluate GenAI's roles in education and how we can use it to promote student-centered learning environments rather than technology-led learning.

Fourth, both students and teachers complained that the AI system could not support social connections among the students and teachers. The feeling of social connections plays an important role in learning, such as promoting a sense of companionship with peers, enabling comparison with others, and improving learning experiences and attainment (Lee & Robbins, 1995; Peacock et al., 2020). The students and teachers wished that the AI system could support social connectedness by allowing students to share their works online and react to others' works by clicking "Like" or leaving comments.

Finally, it was perceived that the current GenAI system does not provide multiple means of communication. This view can be connected to the systems' lack of accessibility (CAST, 2018). The lack of offering multiple means of communication, such as speech-to-text, can be a big barrier for students with limited ability for activities like typing. To ensure the accessibility and inclusivity of the technology for diverse learners, it is essential to offer multiple means of communication between students and the AI system (CAST, 2018; Song et al., 2024b).

## V. Discussion

This study found many educational benefits and challenges of creative mathematical writing with GenAI, as perceived by students and teachers. The identified benefits widely cover students' development of cognitive and affective learning abilities, such as creative thinking skills, mathematical knowledge, AI literacy, and affective domain. Notably, despite the public concerns of GenAI being a barrier to students' creativity (Shidiq, 2023), this study found that students perceived GenAI as helping foster aspects of their creativity, ranging from divergent to flexible thinking (Theme #1). This finding aligns with a body of previous literature that revealed positive effects of GenAI in students' creative writing of algebra (Nouari et al., 2024), science fiction (Wafa et al., 2024), or foreign language writing (Weng & Chiu, 2023). Similarly, writing with GenAI was perceived to help foster students' domain knowledge, which, in this study, is mathematical knowledge (Theme #2). This finding aligns with the previous studies that found an improvement in students' mathematical language use and knowledge after co-creating stories with GenAI (Zhang et al., 2024). This study's finding disentangles the mechanism of how the learning experience with [ANONYMOUS] helps foster students' mathematical understanding. Furthermore, we found that working with GenAI promoted students' AI literacy (Theme #3). AI literacy has become an important competency among K-12 learners, and recent educational efforts have taken the approach of teaching AI as learning content to foster AI literacy (Song et al., 2023). However, this study reveals that exposure and hands-on experience with cutting-edge GenAI technology can potentially develop students' AI literacy. Lastly, while many previous studies in the field of AI-assisted learning focused on the cognitive outcomes of their intervention (Ng et al., 2022; Zhang et al., 2024), this study suggests that creative writing with GenAI also offers opportunities to support students affective domain towards learning (Theme #4), enhancing highly human abilities such as problem-solving skills and diverse areas of affective domain in learning (e.g., having interests, curiosity, and ownership in learning).

Although GenAI offers immense potential to transform education, the desired educational outcomes do not occur merely by using advanced AI computing technologies (Castañeda & Selwyn, 2018; Kim & Cho, 2023). Identified themes regarding the challenges of GenAI support emphasize the importance of deliberate design of instruction, curriculum, technology interface, and learning environment. First, we learned that some students lack competencies in critically evaluating AI-generated content and generating their own original ideas for math problems (Theme #5). This suggests that AI-integrated learning technologies should be developed to complement students' lack of competencies and equipped with diverse pedagogical support. For example, AI-integrated learning technologies should consider pedagogical skills, such as adaptive and personalized feedback, which should be implemented to promote students' autonomous learning and social connections. To be like that, AI should be developed to be more aware of human development and seen as a "collaborator" or "team member" to complement human abilities and work synergistically with humans (Kim et al., 2024b; Kim & Cho, 2023).

On the other hand, our finding also suggests many challenges of AI technologies (Theme #6). One notable finding is that students did not like AI to be "too perfect." Unlike other fields of AI, such as medical AI or AI for business, where the AI's outcomes have a consequential impact on people's lives, education is a special context; AI in education intends to provide support for students' cognitive and affective development, rather than merely providing the best quality outcomes or high accuracy (Barrett & Pack, 2023). In this AI-assisted learning environment, AI's "perfect" performance is not always helpful. Sometimes, AI could intentionally perform imperfections and present room for improvement to allow students to play an important role in this improvement and thereby learn. For example, AI could be developed to simulate a "not-so-knowledgeable" peer to promote students' learning by teaching (Song et al., 2024a). Therefore, adjusting AI's performances to best serve educational purposes is important.

## VI. Conclusion

This study illustrates the students' and teachers' perceived benefits and challenges of creative mathematical writing with GenAI. We collected interview data, survey responses, and classroom observations based on middle school classroom studies with 79 5th-grade students and 6 teachers. We yielded six themes and nineteen subthemes through a thematic analysis of the collected interviews, triangulated with open-ended survey responses and classroom observation. The identified benefits suggest the potential of GenAI in mathematical creative writing, which can be generally applied in other related pedagogical areas, such as mathematical problem-solving, story-based math learning, and creative writing. On the other hand, the suggested barriers and challenges provide important design implications for the GenAI-powered pedagogical tools or curriculum development in the context of mathematical creative writing. This study contributes to the body of AI in Education research by investigating the lived experiences of GenAI-supported learning and the subjective viewpoints of students and teachers from the actual classroom, delivering their voice to the researchers and developers of AI in Education.